# Human-AI collaboration for modeling heat conduction in nanostructures

Wenyang Ding[1#], Jiang Guo[1,3#], Meng An[1,2], Koji Tsuda[3,4], Junichiro Shiomi[1,2,4*]

[1]Department of Mechanical Engineering, The University of Tokyo, Japan

[2]Institute of Engineering Innovation, The University of Tokyo, Japan

[3]Department of Computational Biology and Medical Science, The University of Tokyo, Japan

[4]RIKEN Center for Advanced Intelligence Project, Japan

[#]These authors contributed equally to this work.

Email: shiomi@photon.t.u-tokyo.ac.jp

**ABSTRACT:** In recent years, materials informatics, which combines data science and artificial intelligence (AI), has garnered significant attention owing to its ability to accelerate material development, reduce costs, and enhance product design. However, despite the widespread use of AI, human involvement is often limited to the initiation and oversight of machine learning processes and rarely includes more substantial roles that capitalize on human intuition or domain expertise. Consequently, true human-AI collaborations, where integrated insights can be maximized, are scarce. This study considers the problem of heat conduction in a two-dimensional nanostructure as a case study. An integrated human-AI collaboration framework is designed and used to construct a model to predict the thermal conductivity. This approach is used to determine the parameters that govern phonon transmission over the full range of frequencies and incidence angles. During operation, the self-learning entropic population annealing technique, which combines entropic sampling with a surrogate machine learning model, generates a small dataset that can be interpreted by a human. Therefore, data-efficient and global modeling is achieved, and parameters with physical interpretations are developed, which can guide nanostructural design to produce materials with specific properties. The proposed framework can leverage the complementary strengths of humans and AI, thereby enhancing the understanding and control of materials.

# Introduction

Significant advancements in artificial intelligence (AI) have been achieved in recent years. Currently, AI models can match and even surpass the abilities of human experts in various domains [1-3]. For example, in clinical trials, AI algorithms have demonstrated the ability to detect referable diabetic retinopathy with accuracy comparable to or greater than that of retinal specialists and trained graders [4]. However, despite the advancements in AI technology, certain weaknesses remain. In particular, AI can struggle with complex decision-making processes where nuanced understanding or emotional intelligence is required. Moreover, AI systems may adapt poorly to dynamic environments, where they can produce unexpected results owing to programming limitations or data biases. These limitations highlight the importance of integrating human knowledge into AI systems. Human-AI collaboration combines the strengths of human intuition and AI computational power. In numerous fields, this combination has proven to be more robust and effective than AI alone [5-7]. In the arts, researchers have suggested that human-AI collaboration can enhance creativity in tasks such as haiku production, where the results surpass those produced by either humans or AI alone [5]. In healthcare, this collaboration can facilitate more empathetic interactions during text-based peer support for mental health [6]. Moreover, in marketing, the ability of AI to analyze vast arrays of customer data, preferences, and behaviors can be combined with human judgment to create personalized and tailored marketing strategies [7].

In science and engineering, data science has been combined with AI in the field of materials informatics to revolutionize material development. This approach has attracted attention because it accelerates innovation, reduces costs, and enhances product design [8-10]. However, despite the widespread use of AI, human involvement is typically limited to setting up and monitoring machine learning processes and it rarely extends to greater engagement that leverages human intuition or expert domain knowledge. Consequently, genuine human-AI collaborations are limited and the potential of integrated processes has not been realized. Therefore, this study aims to design an integrated collaborative framework with enhanced human-AI interaction in the field of materials informatics. The proposed framework is designed to combine the strengths of humans and AI to increase innovation and the efficacy of materials research and development.

Heat conduction in two-dimensional (2D) van der Waals (vdW) heterostructures is chosen as a case study. The thermal characteristics of materials must be optimized to improve their energy conversion efficiency, thermal insulation capabilities, and heat dissipation, and to facilitate the creation of multifunctional materials that can drive technological advancements across various domains [9, 11, 12]. There, characterizing transport of phonons, the quantized vibrational modes inherent to the crystalline lattices of solids, is important. VdW heterostructures, created by precisely layering diverse 2D materials [13], exhibit remarkable thermal capabilities [14, 15] and offer an optimal foundation for exploring coherent phonon transport because of their seamlessly integrated interfaces [16].

Graphene is the first 2D material to be developed, and it is important owing to its excellent properties, which include high electrical conductivity, mechanical strength, and flexibility [17, 18]. Recently, transition metal dichalcogenide (TMDC) materials, which have the chemical formula $MX_2$ (where M = Mo or W and X = S, Se, or Te), have attracted attention owing to their excellent thermal and electrical properties [19]. In particular, $WS_2$ is a good candidate for thermoelectric applications because it has a high Seebeck coefficient [20, 21] and comparatively low thermal conductivity [22, 23]. Therefore, graphene-$WS_2$ heterostructures have gained prominence in advanced materials research for combining the good mechanical strength and flexibility of graphene with the superior thermal properties of $WS_2$.

In human-AI collaboration, the design of work structures can significantly affect the integration of AI within teams and the success of collaborative efforts [24]. Collaborations can be structured in various ways to optimize interactions between humans and AI systems [25]. This study combines the strengths of AI and human expertise by utilizing two models with sequential structures: AI-first human-aided (AH) and human-first AI-enhanced (HA). In the AH model, AI is used to process the initial data, and then the insights are passed to humans for further analysis and refinement. By contrast, in the HA model, human experts evaluate and interpret the data based on their unique insights, and then the processed information is passed to the AI for enhancement. The proposed workflow integrates these approaches into an AI-human-AI sequence, which maximizes the synergistic potential of human and machine intelligence.

In this work, the first AI combines the Self-Learning Entropic Population Annealing (SLEPA) algorithm, which we developed in a previous study [26], with the mode-resolved atomistic Green's function (AGF) to generate a small dataset that mimics the original large dataset (as shown in Figure 1, Step 1). The unbiased nature of SLEPA allows it to reveal the full scope of complex systems with minimal computational resources and facilitates efficient identification and selection of optimal values. Owing to these capabilities, a comprehensive understanding of the thermal conductivity distribution in heterostructures can be obtained by analyzing a small subset of candidates. Thus, the interpretation of complex datasets by humans can be improved, which enables the identification of meaningful features and insights that would not be revealed by purely data-driven frameworks (Figure 1, Step 2). The mechanism underlying these features is analyzed using the mode-resolved AGF, which provides a comprehensive understanding of the thermal conductivity contributions of coherent phonons. The second AI utilizes additional machine learning techniques to construct a model with good interpretability and accuracy (Figure 1, Step 3). In summary, the proposed human-AI collaboration framework includes machine learning−SLEPA, human analysis, and further machine learning−Random Forest (RF) and Symbolic Regression (SR) algorithms, which allows the development of a robust and interpretable predictive model for vdW heterostructures.

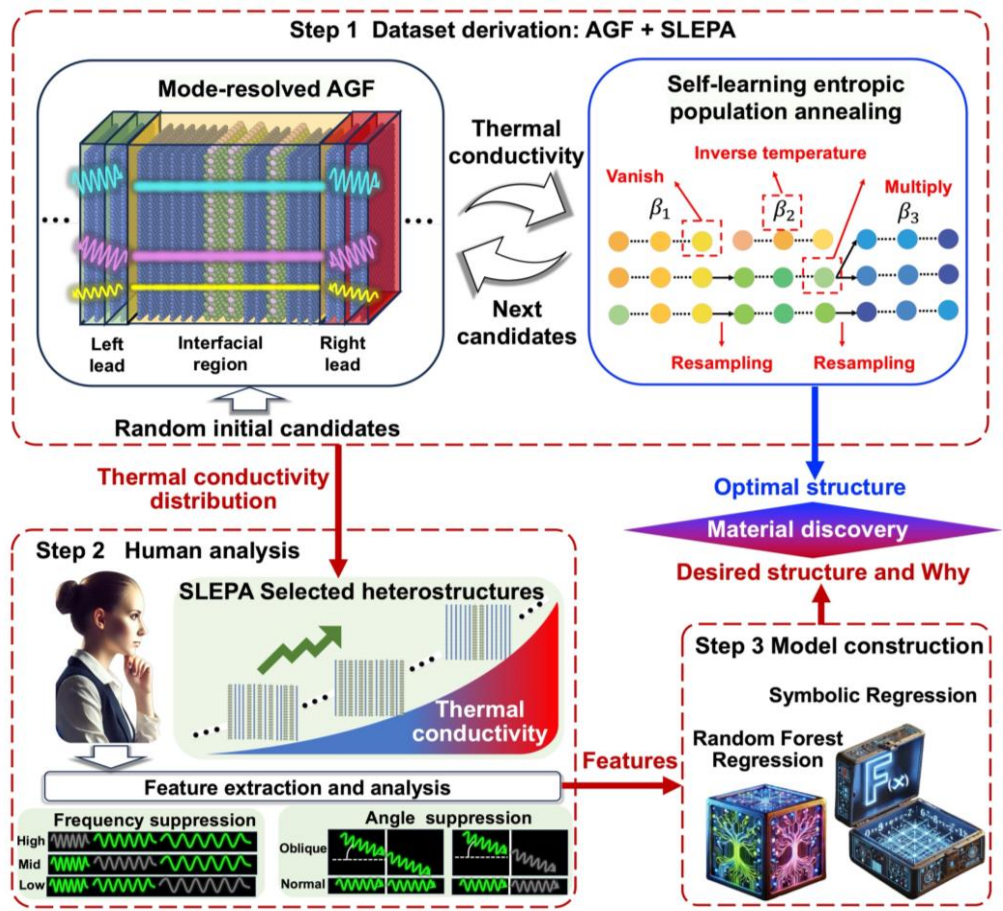

Figure 1. Schematic of human-AI collaboration.

# Results

**Database generation.** Graphene-WS$_2$ heterostructures were assembled through stochastic interfacial arrangements of graphene and WS$_2$ layers (see Methods for details). A binary array representation (descriptor) was used to describe each potential heterostructure. The graphene and WS$_2$ layers were labeled "0" and "1," respectively. The number of layers $N$ was fixed at 14. Therefore, the number of potential candidates $2^N$ was 16 384.

**Validation of SLEPA using ten-layer graphene-WS$_2$ heterostructures.** Before investigating human-AI collaboration, the effectiveness of SLEPA must be confirmed. Therefore, the results obtained using SLEPA, Bayesian optimization, and random sampling were compared to the ground truth. The stacking number was set to ten, and the thermal conductivities of all possible candidates ($2^{10}$ = 1024) were calculated, as illustrated in Figure 2(a). The thermal conductivity distributions for 100 cases (10%) were plotted for each method to enable visual comparison, as illustrated in Figure 2. The *x*-axis shows the decimal number equivalent to the binary number representing the stacking order of the heterostructure. For example, the binary number for a superlattice is "0101010101," which corresponds to the decimal "170."

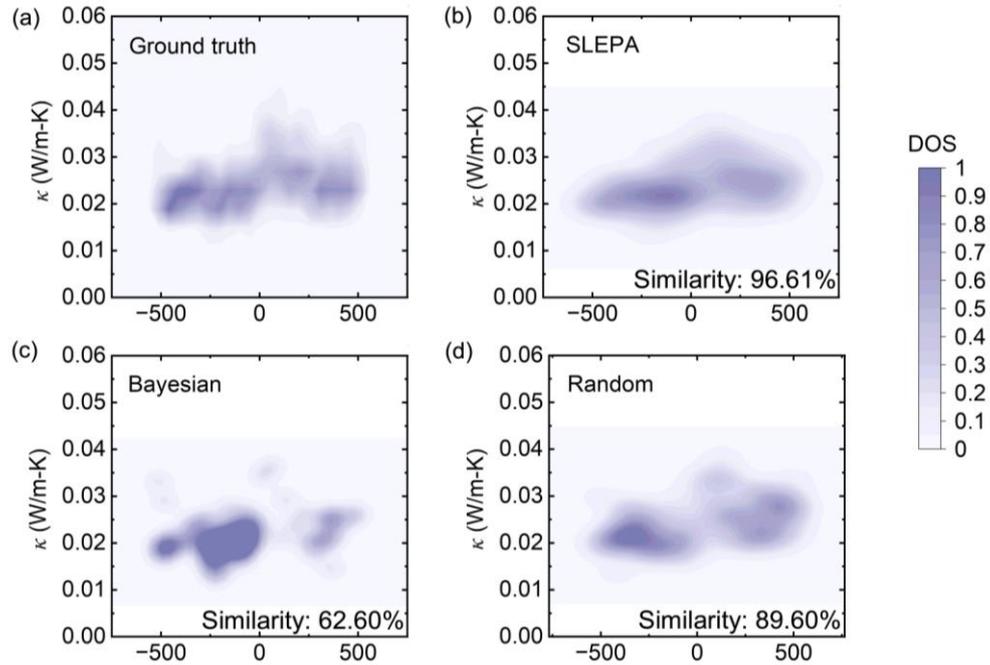

Figure 2. Distribution of thermal conductivity. Results for (a) ground truth, (b) SLEPA, (c) Bayesian optimization, and (d) random sampling. Density of states (DOS) illustrates the distribution of thermal conductivity across the space where the stacking order acts as the variable.

Color histograms were used to quantify the similarity between images, as shown in Figure 2. The thermal conductivity distribution of SLEPA (Figure 2(b)) resembled that of the ground truth (Figure 2(a)), which demonstrates the effectiveness of SLEPA. By contrast, the distribution of the Bayesian optimization (Figure 2(c)) was concentrated around the left local minimum, which demonstrates its limited ability to accurately represent the true distribution. Random sampling (Figure 2(d)) achieved suboptimal results because it produced a scattered and haphazard thermal conductivity distribution with several isolated and concentrated spots. These spots indicate that correct predictions were occasionally achieved by chance; however, this method did not match the consistency and precision achieved by the ground truth and SLEPA methods. Therefore, this analysis confirms that the proposed methodology is robust and constitutes a reliable foundation for subsequent investigations The thermal conductivity distributions obtained using SLEPA, Bayesian optimization, and random sampling with 20%, 30%, and 40% of the total candidates are illustrated in Figures S1, S2, and S3, respectively. These figures show that the percentage of the total candidates used for the calculations did not affect the superior performance of SLEPA.

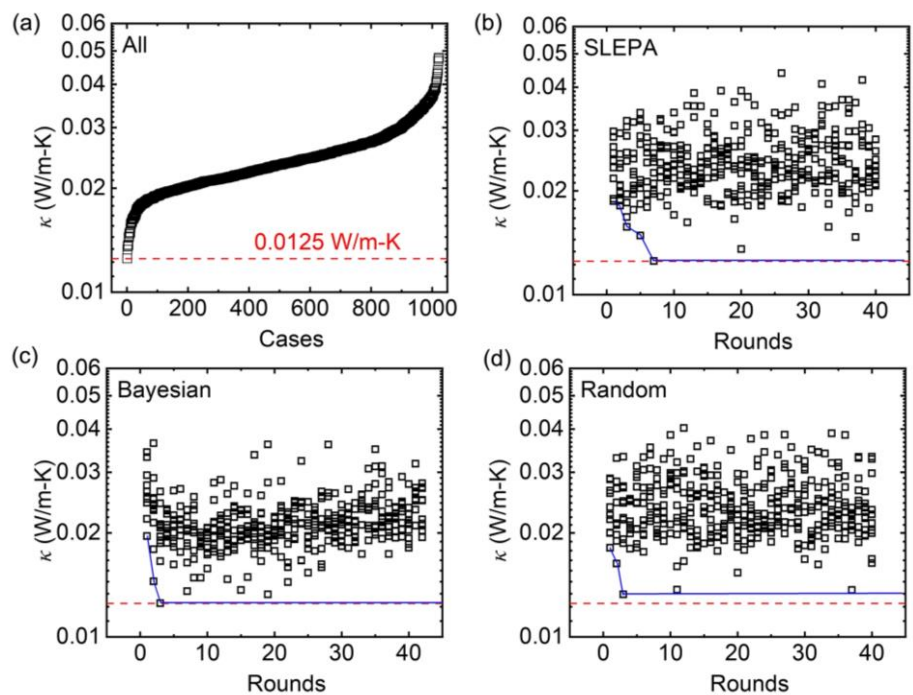

Figure 3. (a) Thermal conductivities for all graphene-$WS_2$ heterostructures. The variation of thermal conductivity for graphene-$WS_2$ heterostructures during the process of (b) SLEPA, (b) Bayesian optimization and (c) Random sampling. The red dash lines represent the real global minimum. The solid blue lines represent the lowest value with the progress of each method. The overlap of red dash lines and blue lines represent the real global minimum can be obtained.

The abilities of SLEPA, Bayesian optimization, and random sampling to obtain the lowest values were also investigated, as shown in Figure 3. SLEPA and Bayesian optimization reached the global minimum after seven and three rounds of calculations, respectively. By contrast, random sampling did not reach the global minimum, even after 40 rounds of calculations. Therefore, although SLEPA was less efficient at determining the global minimum than Bayesian optimization, both methods reached the global minimum within 10% of the maximum number of calculations.

SLEPA can simultaneously obtain the thermal conductivity distribution and identify the optimal heterostructure. This can help experts to analyze the mechanisms that control the thermal conductivity of heterostructures. Specifically, the thermal conductivity distribution can be used as a trial-and-error pool, whereas the identified optimal heterostructure provides insight that can guide efforts to produce materials with lower thermal conductivity.

**SLEPA for 14-layer graphene-WS$_2$ heterostructures.** After validation, SLEPA was applied to 14-layer graphene-WS$_2$ heterostructures with 10% of the total number of candidates. The variations in the thermal conductivity during the application of SLEPA are shown in Figure 4(a). The dashed red line indicates the progress of the lowest thermal conductivity throughout the SLEPA iterations. The optimized heterostructure, represented by the sequence "11000000101101," was identified after 13 rounds (1300 cases) of SLEPA, and corresponded to a thermal conductivity of 0.018 W/m-K. In addition to the optimal heterostructure with ultralow thermal conductivity, SLEPA also outputs the thermal conductivity distribution, as illustrated in Figure 4(b).

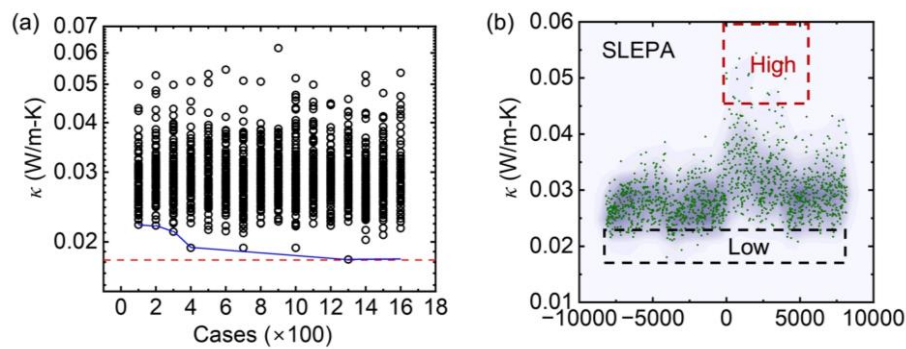

Figure 4. (a) Variations in the thermal conductivity of graphene-WS$_2$ heterostructures during the application of SLEPA. (b) Thermal conductivity distribution obtained using SLEPA.

**Identification of meaningful features.** As illustrated in Figure 1, Step 2, human experts extract features from datasets obtained using a combination of SLEPA and mode-resolved AGF. During this process, the small dataset obtained for the thermal conductivity distribution was used as the basis for extracting and testing features, and the optimized heterostructure was used as inspiration to determine possible features. A hands-on empirical research methodology was used to identify and validate significant features that affect thermal conductivity. This approach involves manual observation and testing of the features to determine their effects on the outcomes. Empirical research methods are well established in scientific investigations as a practical approach to hypothesis testing and model validation through direct observation and experimentation [27].

Figure 5(a) illustrates the two heterostructures with the lowest thermal conductivities from the region with low thermal conductivity (indicated by the dashed black frame in Figure 4(b)), and the two heterostructures with the highest thermal conductivities from the region with high thermal conductivity (indicated by the dashed red frame in Figure 4(b). In the heterostructures with low thermal conductivities, the layers at both ends were composed of $WS_2$. By contrast, in the heterostructures with high thermal conductivities, the layers at both ends were composed of graphene. This indicates that the end layers of heterostructures with low thermal conductivity have approximately zero probability of being graphene, as illustrated in Figure 5(b). Twenty heterostructures from each of the regions with low and high thermal conductivity were selected for further analysis to determine whether this pattern held for a broader sample. The probability of each layer being graphene was calculated for the samples from each region, as illustrated in Figure 5(c). For heterostructures with low thermal conductivities, the end layers were usually $WS_2$, whereas the middle layers were usually graphene. For the heterostructures with high thermal conductivities, this arrangement was reversed.

This phenomenon is quantified using parameter $P_a$, defined as:

$$P_a = \frac{n_0 + 1}{sum_0 + 1}, \tag{1}$$

where $sum_0$ denotes the total number of graphene layers in the heterostructure and $n_0$ denotes the number of graphene layers between the outermost $WS_2$ layers. The effectiveness of the parameter was tested by visualizing the values of $P_a$ for heterostructures from the regions with low and high thermal conductivity, as shown in

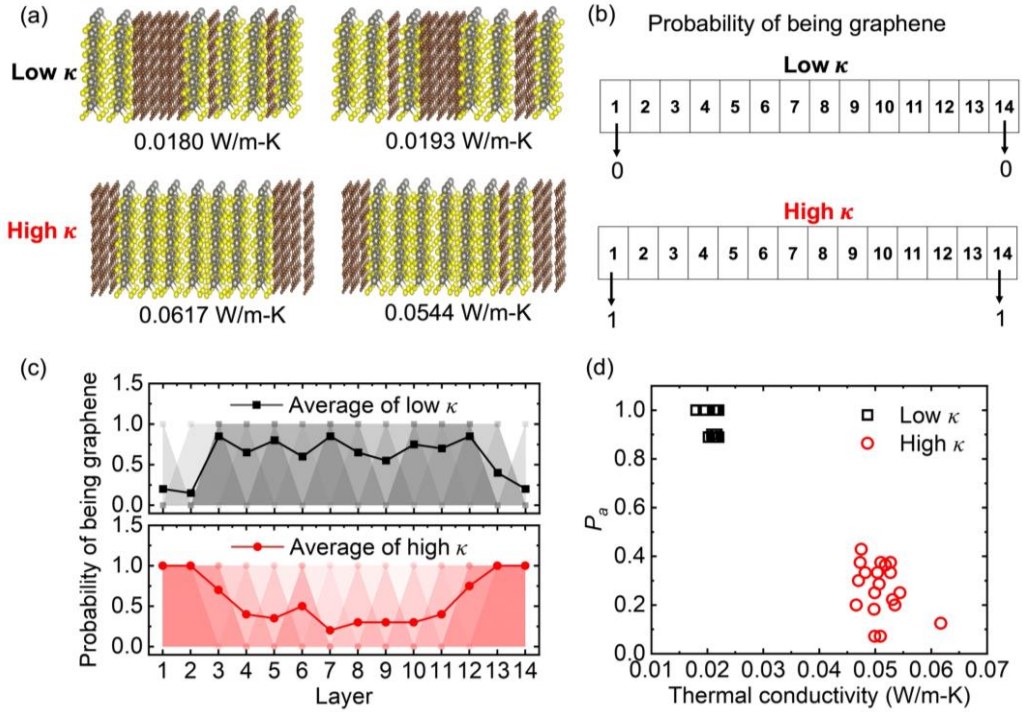

Figure 5. (a) Graphene-WS$_2$ heterostructures, two with low and two with high thermal conductivities. (b) Probabilities of the end layers being graphene for heterostructures with low and high thermal conductivities (two each). (c) The smaller dots with shadow represent the probability of each layer being graphene or heterostructures from regions with low and high thermal conductivities (20 each). The solid lines with dots indicate average probability of each layer being graphene for heterostructures from regions with low and high thermal conductivities (20 each). (d) Values of parameter $P_a$ for heterostructures from regions with low and high thermal conductivities.

Figure 5(d). The heterostructures from the region with low thermal conductivity possessed high values of $P_a$, whereas those from the region with high thermal conductivity possessed relatively low values of $P_a$. This indicates that $P_a$ can distinguish between heterostructures with very high and very low thermal conductivities based on the stacking order.

The values of $P_a$ for all the candidates identified by SLEPA are illustrated in Figure 6(a). The results indicate that heterostructures with larger values of $P_a$ typically have lower thermal conductivities. Therefore, $P_a$ is an effective descriptor that can be used to determine the thermal conductivity of heterostructures. However, several heterostructures with the same $P_a$ values possess different thermal conductivities. Therefore, additional descriptors are required to distinguish between heterostructures with the same $P_a$. Three groups of heterostructures with $P_a$ values of 0.444, 0.625, and 0.800 were selected. Based on observations of the heterostructures in each group, we hypothesized that the subsequence length of the graphene layers may be an important factor affecting thermal conductivity. A subsequence is a small sequence within each heterostructure, such as "101," "1001," and "10001." The subsequence length is equal to the number of graphene layers within the subsequence. For example, subsequences "101" and "1001" have subsequence lengths $n$ of 1 and 2, respectively. Each heterostructure contains several subsequences; therefore, we defined "number of graphene" to represent the summation of the subsequence lengths across all the subsequences in a heterostructure. Hence, we verified our hypothesis by investigating the relationship between the thermal conductivity and the number of graphene layers in the heterostructures, as illustrated in Figure 6(b). For $n = 1$, the thermal conductivity increased as the number of graphene increased. For $n \geq 2$, the thermal conductivity decreased as the number of graphene increased. This indicates that heterostructures with subsequence lengths greater than two tend to have lower thermal conductivities. Therefore, parameter $P_b$ is defined as:

$$P_b = \frac{n_{>00} + 1}{sum_0 + 1}, \tag{2}$$

where $sum_0$ denotes the total number of graphene layers in the heterostructure and $n_{>00}$ denotes the number of graphene layers in sequences that contain at least two consecutive graphene layers. The effectiveness of $P_b$ was verified by investigating the relationship between $P_b$ and thermal conductivity, as shown in Figure 6(c). Evidently, $P_b$ is an effective descriptor that can be used to determine the thermal conductivity of heterostructures.

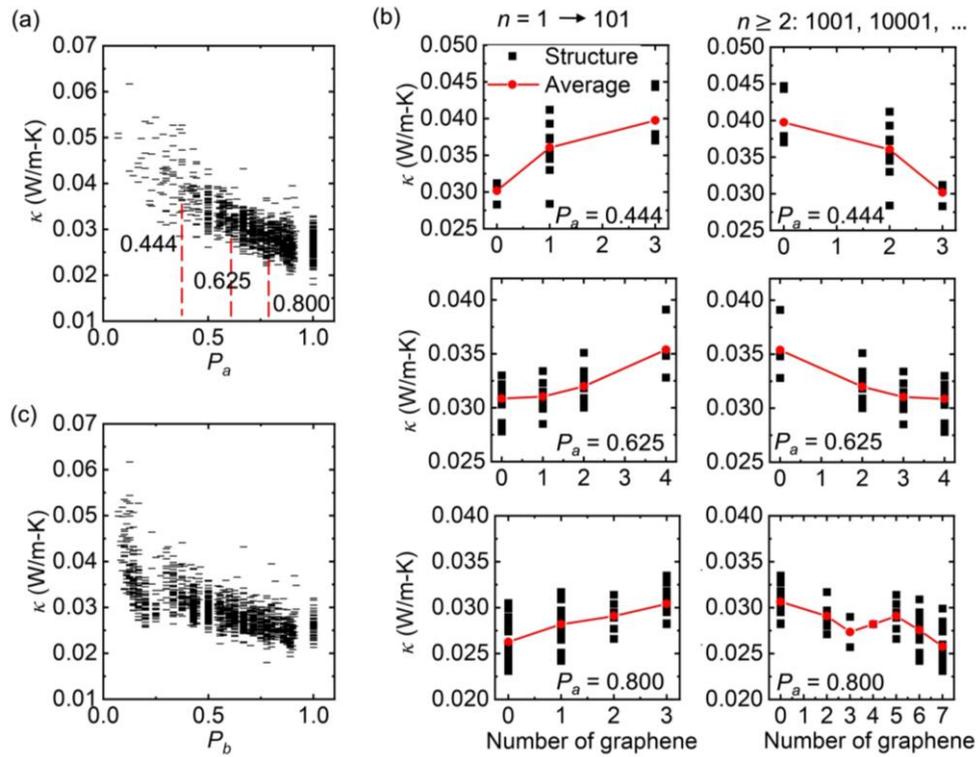

Figure 6. (a) Relationship between the thermal conductivities of the heterostructures identified by SLEPA and parameter $P_a$. (b) Relationship between thermal conductivity and subsequence length of graphene. Here, $n$ is the subsequence length and $n = 1, 2, 3…$ represents subsequences "101", "1001", "10001," etc. The $x$-axis, titled "number of graphene," denotes the summation of the graphene subsequence length across all the subsequences. (c) Relationship between the thermal conductivity of the heterostructures identified by SLEPA and parameter $P_b$.

The parameter $P_b$ is related to the graphene subsequence length; therefore, another parameter may be derived from the $WS_2$ subsequence length. Two groups of heterostructures with different $P_b$ values were selected for analysis. As for the definition of the graphene subsequence length, the $WS_2$ subsequence length is equal to the number of $WS_2$ layers in the subsequence. For example, subsequences "010" and "0110" have subsequence lengths $n$ of 1 and 2, respectively. We also defined "number of $WS_2$" as the summation of the $WS_2$ subsequence lengths across all the subsequences in a heterostructure. The relationship between the thermal conductivity and the number of $WS_2$ in the heterostructure is illustrated in Figure 7(a). For all the analyzed heterostructures, no significant relationship was observed.

Therefore, the deadlock was broken by observing the optimal heterostructures identified by SLEPA, as shown in Figure 7(b). Subsequences with lengths of one or two coexisted in the optimal heterostructure. Therefore, inspired by the optimal heterostructure, we multiplied $n = 1$ by $n = 2$ and observed the relationship with the thermal conductivity, as shown in Figure 7(c). As the value of $n = 1$ multiplied by $n = 2$ increased, the thermal conductivity decreased. Therefore, parameter $P_c$ is defined as:

$$P_c = \frac{n_1 \times n_{11} + 1}{sum_1 + 1}, \tag{3}$$

where $sum_1$ denotes the total number of $WS_2$ layers, $n_1$ denotes the number of $WS_2$ layers in sequences containing only one $WS_2$ layer, and $n_{11}$ denotes the number of $WS_2$ layers in sequences containing two consecutive $WS_2$ layers. Figure 7(d) shows the relationship between the thermal conductivities of the heterostructures identified by SLEPA and parameter $P_c$. Although parameter $P_c$ is not as effective as parameters $P_a$ and $P_b$, it still has a significant effect on the maximum thermal conductivity of a heterostructure.

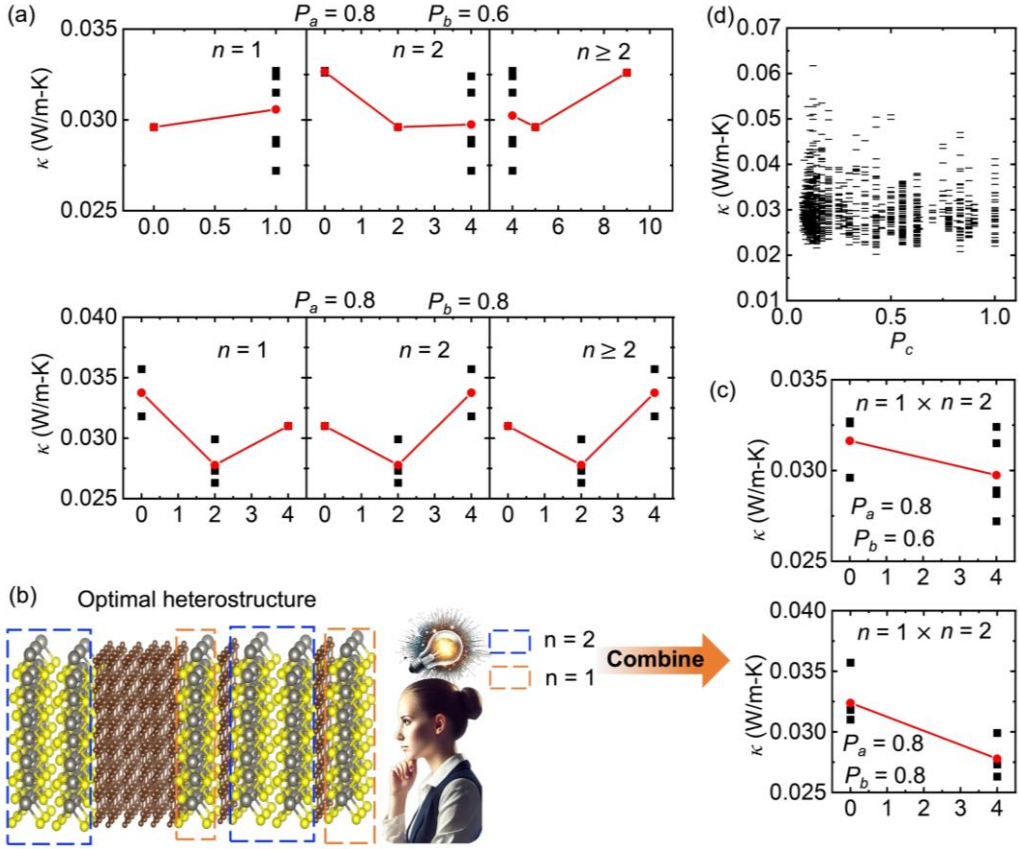

Figure 7. Relationship between thermal conductivity and subsequence length of WS$_2$. (a) $P_a = 0.8$, $P_b = 0.6$ and $P_a = 0.8$, $P_b = 0.8$. Here, $n$ is subsequence length and $n = 1, 2$ represents subsequences "010" and "0110," and $n \geq 2$ represents subsequences "0110," "01110," etc. The $x$-axis, titled "number of WS$_2$," denotes the summation of the WS$_2$ subsequence length across all the subsequences. (b) Optimal heterostructure identified by SLEPA. Blue and orange frames represent WS$_2$ subsequence lengths of 1 and 2, respectively. (c) Relationship between thermal conductivity and number of WS$_2$, derived from $n = 1$ multiplied by $n = 2$. (d) Relationship between the thermal conductivities of heterostructures identified by SLEPA and parameter $P_c$.

**Construction of a predictive model.** The relationships between the identified features must be determined to construct a predictive model. This was achieved by returning to machine learning techniques. An RF model was trained using the three parameters $P_a$, $P_b$, and $P_c$. The RF algorithm is an ensemble learning method that constructs multiple decision trees during training and aggregates their predictions to improve the overall accuracy [28]. This method allows us to utilize the collective predictive power of numerous decision trees, which produces a highly accurate model for predicting thermal conductivity. The RF model, known for its robustness and accuracy, proved the effectiveness of the parameters. The actual thermal conductivities were compared to the values predicted by the trained RF model, as illustrated in Figure 8(a). This demonstrates that the constructed model has high accuracy, with an $R^2$ value of 0.70.

Although the RF model is highly accurate, it lacks interpretability and functions like a closed box, as shown in Figure 1, Step 3. Therefore, we incorporated SR into the model, which searches for the mathematical expression that best fits the relationship between the features and the target variable [29]. This approach provided transparency and insight that are not typically available for black-box models such as RF models. The final model is expressed as:

$$\kappa = \frac{0.178}{9.89 + P_a + P_b + 0.143 P_c} + 0.0109. \tag{4}$$

The actual thermal conductivities were compared to the predicted values obtained from the trained SR model, as shown in Figure 8(b). The accuracy of the constructed model was 64%, which was only slightly lower than that of the RF model. Moreover, the model, which was developed using the identified features, transforms raw data into insightful information that facilitates a deeper understanding of the underlying mechanisms.

According to the model, the theoretical minimum thermal conductivity is 0.0256 W/m-K, and the corresponding heterostructure is "10000100110011," as shown in the inset of Figure 8(c). The actual thermal conductivity of this heterostructure is 0.0180 W/m-K, which is only slightly higher than that of the optimal heterostructure, as shown in Figure 7. This confirms the effectiveness of SLEPA in optimization. Moreover, the features were identified by experts, whose accurate data analysis was facilitated by the fair distribution provided by SLEPA. The unbiased nature of SLEPA provided the experts with a comprehensive picture, avoided skewed representation, and ensured that the data were representative of the entire parameter space. These characteristics ensure that the resulting model and mechanisms are accurate and generalizable; hence, robust and broadly applicable insights can be obtained across different contexts and scenarios.

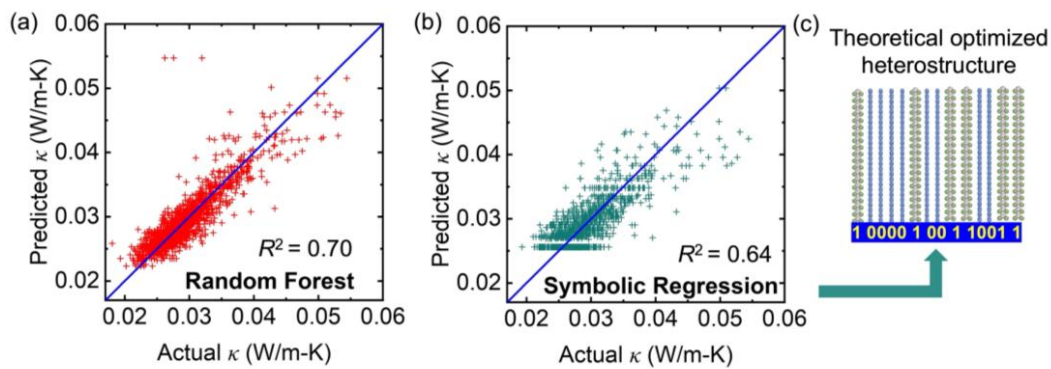

Figure 8. Comparison between actual and predicted thermal conductivities obtained from the (a) trained RF model, (b) trained SR model, and (c) theoretical optimized heterostructure according to the SR model.

**Mechanism under the three features.** Finally, the mechanisms underlying the three parameters were investigated. Two heterostructures with the same values of $P_b$ and $P_c$ and different values of $P_a$ were selected to investigate $P_a$. The heterostructures were "00000100100000" and "10000000000001," and they were labeled *Stru. $a_1$* and *Stru. $a_2$*, respectively. The parameters corresponding to these heterostructures are listed in Table S1 and their thermal conductivities are shown in Figure 9(a). The $P_a$ of *Stru. $a_2$* was larger than that of *Stru. $a_1$*, and it exhibited a lower thermal conductivity. To understand the underlying mechanisms, the relationship between the spectral thermal transmission of the heterostructures and the frequency was investigated, as shown in Figure 9(b). The results show that the low-frequency phonons (0–1.2 THz) were suppressed in *Stru. $a_2$*. The relationships between the normally and obliquely incident phonons and the frequency were also investigated, as shown in Figures 9(c) and (d). This shows that only the normally incident phonons in *Stru. $a_2$* are suppressed significantly.

Two heterostructures with the same values of $P_a$ and $P_c$ and different values of $P_b$ were selected to investigate $P_b$. The heterostructures were "01010101010101" and "01100011000111," and they were labeled *Stru. $b_1$* and *Stru. $b_2$*, respectively. The parameters corresponding to these heterostructures are listed in Table S2 and their thermal conductivities are shown in Figure 9(e). The $P_b$ of *Stru. $b_2$* was larger than that of *Stru. $b_1$*, and it exhibited a lower thermal conductivity. To understand the underlying mechanisms, the relationship between the spectral thermal transmission of the heterostructures and the frequency was investigated, as shown in Figure 9(f). The results show that the middle-frequency phonons (1.0–2.3 THz) were suppressed in *Stru. $b_2$*. The relationships between the normally and obliquely incident phonons and the frequency were also investigated, as shown in Figures 9(g) and (h). This shows that only the normally incident phonons in *Stru. $b_2$* are suppressed significantly.

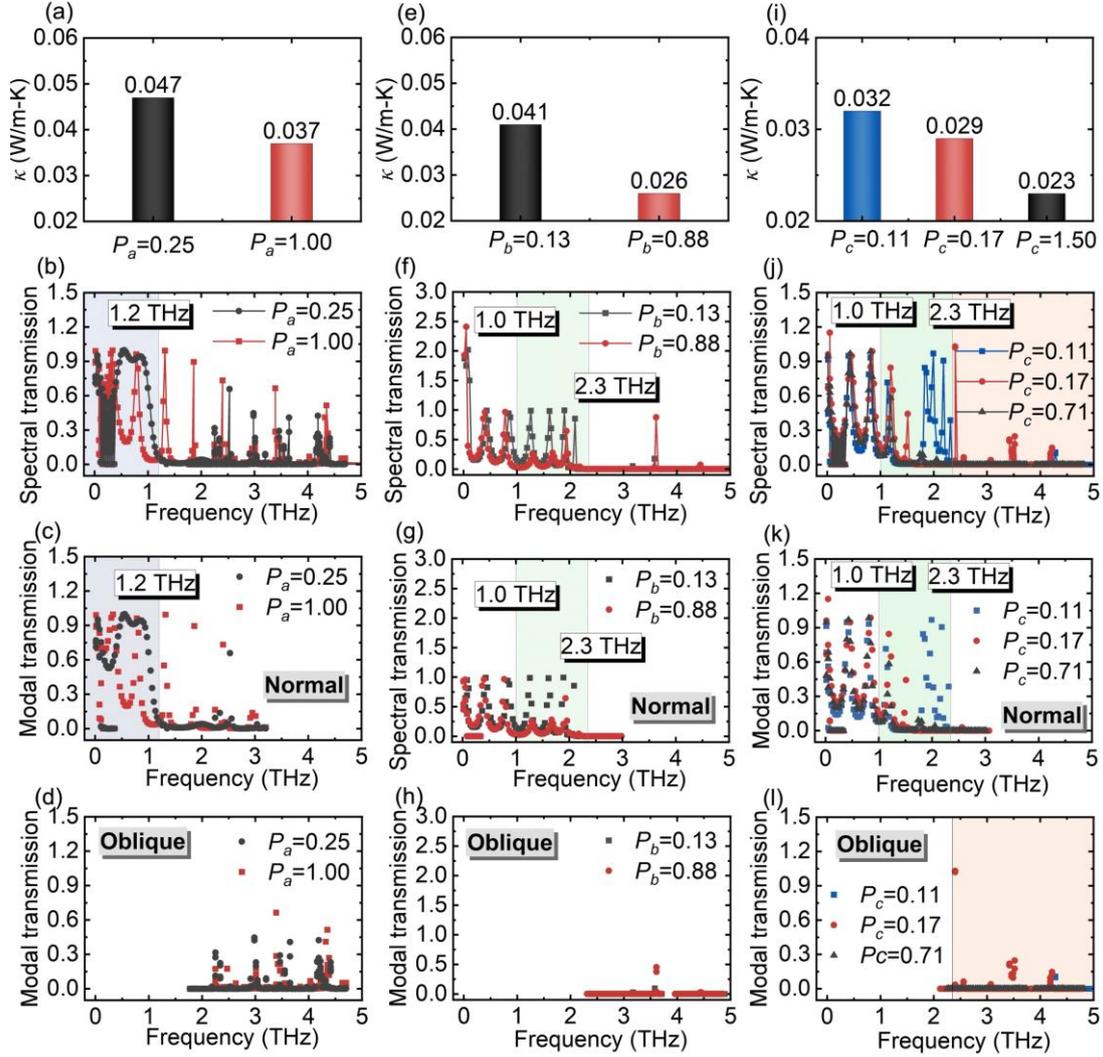

Figure 9. Thermal conductivities and frequency-dependent properties of (a–d) *Stru. a₁* and *Stru. a₂*, (e–h) *Stru. b₁* and *Stru. b₂*, and (i–l) *Stru. c₁*, *Stru. c₂*, and *Stru. c₃*. (a), (e), (i) Thermal conductivities. (b), (f), (j) Spectral transmission. (c), (g), (k) Normally incident phonon modal transmission. (d), (h), (l) Obliquely incident phonon modal transmission.

Three heterostructures with the same values of $P_a$ and $P_b$ and different values of $P_c$ were selected to investigate $P_c$. The heterostructures were "11001100110011" (which contains only "11" WS$_2$ sequences), "10010010010001" (which contains only "1" WS$_2$ sequences), and "11001100010001" (which contains both "11" and "1" WS$_2$ sequences), and they were labeled *Stru. c₁*, *Stru. c₂*, and *Stru. c₃*, respectively. The parameters corresponding to these heterostructures are listed in Table S3 and their thermal conductivities are shown in Figure 9(i). The $P_c$ of *Stru. c₃* was larger than those of *Stru. c₁* and *Stru. c₂*, and it exhibited a lower thermal conductivity. To understand the underlying mechanisms, the relationship between the spectral thermal transmission of the heterostructures and the frequency was investigated, as shown in Figure 9(j). The results show that the middle-frequency (1.0–2.3 THz) and high-frequency (2.3–5.0 THz) phonons were suppressed in *Stru. c₃*. The relationships between the normally and obliquely incident phonons and the frequency were also investigated, as shown in Figures 9(k) and (l). This shows that normally incident phonons at middle frequencies and obliquely incident phonons at high frequencies are suppressed.

The frequencies and incidence angles where phonon suppression occurs are listed in Table 1. In summary, $P_a$ suppresses normally incident phonon modal transmission at low frequencies, $P_b$ suppresses normally incident phonon modal transmission at middle frequencies, and $P_c$ suppresses normally incident phonon modal transmission at middle frequencies, and obliquely incident phonon modal transmission at high frequencies.

These parameters provide clear and actionable steps, which form a practical and straightforward framework for implementation: (1) ensure that the heterostructure begins and ends with $WS_2$ layers, (2) favor multilayer graphene configurations, (3) limit $WS_2$ to one or two contiguous layers, and (4) aim for a balanced combination of graphene and $WS_2$. These principles can help researchers navigate the intricate terrain of material design, where direct replication of simulations is often challenging. This approach simplifies the experimental process and increases the reproducibility and applicability of the findings. Therefore, these principles go beyond methodological tools and help to bridge the gap between theoretical simulations and experimental applications, thereby advancing scientific understanding and the practical utility of research.

Table 1. Range of frequences and incidence angles corresponding to phonon suppression for $P_a$, $P_b$, and $P_c$.

|  | Low frequency | Mid frequency | High frequency |
| --- | --- | --- | --- |
| Normally incident | $P_a$ | $P_b$, $P_c$ | No mode |
| Oblique incident | No mode | No mode | $P_c$ |

## Conclusions

Our investigation of graphene-$WS_2$ heterostructures demonstrated the efficacy of combining advanced machine learning techniques with human expertise to model and predict thermal properties. SLEPA was combined with mode-resolved AGF analysis, which allowed us to identify and interpret meaningful patterns in phonon transport. The derived parameters, $P_a$, $P_b$, and $P_c$, have a significant effect on thermal conductivity, and each parameter contributes to phonon suppression at specific frequencies and incidence angles. The final SR model based on these parameters provides a robust and interpretable framework that can be used to predict the thermal conductivities of graphene-$WS_2$ heterostructures. The successful application of SLEPA, SR, and human-AI collaboration methodologies highlights the potential of integrating advanced machine learning techniques with human expertise in the field of materials science. By bridging the theoretical and practical domains, our findings pave the way for the design and optimization of 2D heterostructures with tailored thermal properties, opening new avenues for efficient energy conversion, thermal insulation, and heat dissipation using advanced materials.

## METHODS

**Parameters of graphene-$WS_2$ heterostructures.** The unit cell parameters for graphene and $WS_2$ were set to 4.260 ×2.460 Å and 5.520 ×3.187 Å, respectively. The heterostructure supercells were created using a 4 ×5 unit cell configuration (17.04 × 12.30 Å) for graphene and a 3 ×4 unit cell configuration (16.56 ×12.75 Å) for $WS_2$.

**SLEPA methodology.** In SLEPA [26], $X$ denotes the set of all possible materials, and $e(x)$ represents the target property. Here, $e(x)$ is referred to as energy, and the acquisition of $e(x)$ is called an observation. Gaussian process regression was adopted as the machine learning model. SLEPA uses $M$ particles and alternates between Markov chain Monte Carlo steps and resampling. The energy $e(x)$ is replaced by the prediction obtained using a machine learning model $\tilde{e}_i(x)$. The model is updated as the number of available training examples increases. The first model $\tilde{e}_1(x)$ is trained using the initial $M$ particles and their energies. The $i$-th model corresponding to $\beta_i$ is trained using the $iM$ examples obtained up to that point. Populations of the samples obtained from SLEPA are selected based on the Boltzmann distributions of the predicted energies. Before applying the multiple histogram method, these populations are corrected by resampling. The weight of the $j$-th particle obtained at $\beta_i$ is given by

$$q_{ij} \propto \exp[-\beta_i(e(x_{ij}) - \tilde{e}_i(x_{ij}))]. \tag{5}$$

In this study, the 100 initial structures were used as the original training dataset and their thermal conductivities were calculated using mode-resolved AGF. Entropic population annealing was then used to predict the next 100 candidate structures from the candidate pool according to their probability of being optimal. A more detailed explanation of SLEPA can be found in Ref. [26].

**Mode-resolved AGF.** Within the AGF framework [30, 31], the system is partitioned into three interconnected components: the left and right semi-infinite leads and the central device region. In this study, the left and right leads consist of two layers of graphene or graphite, and the central region consists of the heterostructure to be analyzed. Subsequently, the LAMMPS molecular dynamic simulator [32] was used to calculate the force constant matrices for the left, right, and central regions. The optimized Tersoff potential [33] was used for the covalent bonds in graphene, the Stillinger-Weber potential [34] was used for the covalent bonds in the $WS_2$ layers, and the Lennard-Jones (L-J) potential was used for the van der Waals interactions between graphene and $WS_2$. The mode-resolved AGF formalism was derived from the work reported by Ong *et al.* [35].

The flux-normalized transmission matrix for phonon transmission from the left to the right lead is

$$t_{RL} = \frac{2i\omega}{\sqrt{a_L a_R}} V_R^{1/2} [U_R^{ret}]^{-1} G^{ret} [U_L^{adv\dagger}]^{-1} V_L^{1/2}, \qquad (6)$$

where $G^{ret} = g_R^{00} H_{RD} G_D^{ret} H_{DL} g_L^{00}$; $a_L$ ($a_R$) is the length of the left (right) sublayer of the lead; $V_L$ ($V_R$) is the projection of the mode group velocity in the left (right) leads along the direction of the temperature gradient, which is zero for evanescent phonons; and $U_R^{ret}$ ($U_L^{adv}$) is a matrix with column vectors that correspond to the extended or decaying evanescent modes of the right (left) leads. The modal transmission coefficient of the $n$th incoming phonon channel in the left lead is $\Xi_{L,n}(\omega) = [\bar{t}_{RL}^\dagger \bar{t}_{RL}]_{nn}$. The spectral phonon transmission can be expressed as the sum of the transmission coefficients at either lead; that is,

$$\Xi(\omega) = \sum_{n=1}^{N_L(+)} \Xi_{L,n}(\omega), \qquad (7)$$

where $N_L(+)$ denotes the number of phonon channels traveling rightward from the left lead. Further details regarding the mode-resolved AGF can be found in the Supplemental Material and Ref. [35].

Finally, the temperature-dependent thermal conductance per unit area $\sigma(T)$ for the device region is calculated using the Landauer formalism [36, 37]:

$$\sigma(T) = \frac{1}{2\pi A} \int_0^{\omega_{max}} \hbar\omega \frac{\partial n}{\partial T}(\omega,T) \Xi(\omega)d\omega, \tag{8}$$

where $A$ is the transverse area, $n(\omega,T)$ is the Bose-Einstein distribution, and $T$ is the temperature.

## Data availability

The datasets generated during and/or analyzed during the current study are available from the corresponding author on reasonable request.

## Acknowledgments

This work was partially funded by CREST (Grant No. JPMJCR21O2) provided by the Japan Science and Technology Agency (JST). The numerical calculations were performed at the Supercomputer Center at the Institute for Solid-State Physics, University of Tokyo, and Masamune-IMR at the Center for Computational Materials Science, Institute for Materials Research, Tohoku University (Project No. 2112SC0507). We used ChatGPT to generate human figures (Figure 1, Step 2) and two AI boxes (Figure 1, Step 3). The funder played no role in study design, data collection, analysis and interpretation of data, or the writing of this manuscript.

## Author information

Authors and Affiliations

**Department of Mechanical Engineering, The University of Tokyo, Japan**

Wenyang Ding, Jiang Guo, Meng An, Junichiro Shiomi

**Institute of Engineering Innovation, The University of Tokyo, Japan**

Meng An, Junichiro Shiomi

**Department of Computational Biology and Medical Science, The University of Tokyo, Japan**

Jiang Guo, Koji Tsuda

**RIKEN Center for Advanced Intelligence Project, Japan**

Koji Tsuda, Junichiro Shiomi

## Contributions

Junichiro Shiomi conceptualized and designed the study. Wenyang Ding conducted the mode-resolved AGF calculations, data analysis, and investigation, and drafted the initial manuscript. Jiang Guo provided supervision on conducting SLEPA, while Meng An and Koji Tsuda contributed to supervision, as well as reviewing and editing the manuscript. All authors actively participated in discussions and contributed to editing the manuscript.

## Corresponding author

Correspondence to Junichiro Shiomi

## Ethics declarations

Competing interests

The authors declare no competing interests.

## References

[1] D. Silver, T. Hubert, J. Schrittwieser, I. Antonoglou, M. Lai, A. Guez, M. Lanctot, L. Sifre, D. Kumaran, T. Graepel, T. Lillicrap, K. Simonyan, D. Hassabis, Science 362, 1140-1144 (2018).

[2] N. Brown, T. Sandholm, Science 365, 885-890 (2019).

[3] K. He, X. Zhang, S. Ren, J. Sun, IEEE International Conference on Computer Vision (ICCV 2015) 1502, (2015).

[4] V. Gulshan, R.P. Rajan, K. Widner, D.J. Wu, P. Wubbels, T. Rhodes, K. Whitehouse, M. Coram, G.S. Corrado, K. Ramasamy, R. Raman, L.H. Peng, D.R. Webster, JAMA ophthalmology DOI (2019).

[5] J. Hitsuwari, Y. Ueda, W. Yun, M. Nomura, Computers in Human Behavior 139, 107502 (2023).

[6] A. Sharma, I.W. Lin, A.S. Miner, D.C. Atkins, T. Althoff, Nature Machine Intelligence 5, 46-57 (2023).

[7] M. Petrescu, A.S. Krishen, Journal of Marketing Analytics 11, 263-274 (2023).

[8] R. Chen, Y. Tian, J. Cao, W. Ren, S. Hu, C. Zeng, Journal of Applied Physics 135, (2024).

[9] S. Hu, S. Ju, C. Shao, J. Guo, B. Xu, M. Ohnishi, J. Shiomi, Mater. Today Phys. 16, 100324 (2021).

[10] S. Ju, T. Shiga, L. Feng, Z. Hou, K. Tsuda, J. Shiomi, Phys. Rev. X 7, 021024 (2017).

[11] J. Spiece, S. Sangtarash, M. Mucientes, A.J. Molina-Mendoza, K. Lulla, T. Mueller, O. Kolosov, H. Sadeghi, C. Evangeli, Nanoscale 14, 2593-2598 (2022).

[12] P.-Z. Jia, Y.-J. Zeng, D. Wu, H. Pan, X.-H. Cao, W.-X. Zhou, Z.-X. Xie, J.-X. Zhang, K.-Q. Chen, J. Phys. Condens. Matter 32, 055302 (2019).

[13] A.K. Geim, I.V. Grigorieva, Nature 499, 419-425 (2013).

[14] L. Zhang, Y. Zhong, X. Qian, Q. Song, J. Zhou, L. Li, L. Guo, G. Chen, E.N. Wang, ACS Applied Materials & Interfaces 13, 46055-46064 (2021).

[15] S. Vaziri, E. Yalon, M. Muñoz Rojo, S.V. Suryavanshi, H. Zhang, C.J. McClellan, C.S. Bailey, K.K.H. Smithe, A.J. Gabourie, V. Chen, S. Deshmukh, L. Bendersky, A.V. Davydov, E. Pop, Sci. Adv. 5, eaax1325 (2019).

[16] R. Guo, Y.-D. Jho, A.J. Minnich, Nanoscale 10, 14432-14440 (2018).

[17] Z.-S. Wu, W. Ren, L. Gao, J. Zhao, Z. Chen, B. Liu, D. Tang, B. Yu, C. Jiang, H.-M. Cheng, ACS Nano 3, 411-417 (2009).

[18] X. Li, J. Yu, S. Wageh, A.A. Al-Ghamdi, J. Xie, Small 12, 6640-6696 (2016).

[19] V. Sorkin, H. Pan, H. Shi, S.Y. Quek, Y.W. Zhang, Crit. Rev. Solid State 39, 319-367 (2014).

[20] G.K. Solanki, D.N. Gujarathi, M.P. Deshpande, D. Lakshminarayana, M.K. Agarwal, Cryst. Res. Technol. 43, 179-185 (2008).

[21] C. Lee, J. Hong, M.-H. Whangbo, J.H. Shim, Chem. Mater. 25, 3745-3752 (2013).

[22] P. Jiang, X. Qian, X. Gu, R. Yang, Adv. Mater. 29, 1701068 (2017).

[23] D.O. Lindroth, P. Erhart, Phys. Rev. B 94, 115205 (2016).

[24] I. Seeber, E. Bittner, R.O. Briggs, T. de Vreede, G.-J. de Vreede, A. Elkins, R. Maier, A.B. Merz, S. Oeste-Reiß, N. Randrup, G. Schwabe, M. Söllner, Information & Management 57, 103174 (2020).

[25] Y. Shrestha, S. Ben-Menahem, G. Krogh, California Management Review 61, 000812561986225 (2019).

[26] J. Li, J. Zhang, R. Tamura, K. Tsuda, Digital Discovery 1, 295-302 (2022).

[27] P. Cohen, *Empirical Methods for Artificial Intelligence*, 1995.

[28] L. Breiman, Mach. Learn. 45, 5-32 (2001).

[29] M. Cranmer, A. Sanchez Gonzalez, P. Battaglia, R. Xu, K. Cranmer, D. Spergel, S. Ho, Adv. Neural Inf. Process. Syst. 33, 17429-17442 (2020).

[30] W. Zhang, T.S. Fisher, N. Mingo, J. Heat Transfer. 129, 483-491 (2006).

[31] W. Zhang, T.S. Fisher, N. Mingo, Numerical Heat Transfer, Part B: Fundamentals 51, 333-349 (2007).

[32] S. Plimpton, J. Comput. Phys. 117, 1-19 (1995).

[33] L. Lindsay, D.A. Broido, Phys. Rev. B 81, 205441 (2010).

[34] A. Mobaraki, A. Kandemir, H. Yapicioglu, O. Gülseren, C. Sevik, Comput. Mater. Sci. 144, 92-98 (2018).

[35] Z.-Y. Ong, Phys. Rev. B 98, 195301 (2018).

[36] L. Yang, B. Latour, A.J. Minnich, Phys. Rev. B 97, 205306 (2018).

[37] A. Dhar, D. Roy, J. Stat. Phys. 125, 801-820 (2006).